\documentstyle[twocolumn,prl,aps,epsfig]{revtex}

\begin{document}
\wideabs{
\title{Ultracold dipolar gases - a challenge for experiments and theory}
\author{M. Baranov$^1$, \L. Dobrek$^1$, K. G\'oral$^{1,2}$, L. Santos$^1$, and M.
Lewenstein$^1$}
\address{(1) Institut f\"ur Theoretische Physik, Universit\"at Hannover, D-30167\\
Hannover, Germany \\
(2) Center for Theoretical Physics and College of Science,\\
Polish Academy of Sciences, Aleja Lotnik\'ow 32/46, 02-668 Warsaw, \\
Poland}
\maketitle

\begin{abstract}
We present a review of recent results concerning the physics of ultracold
trapped dipolar gases. In particular, we discuss the Bose-Einstein
condensation for dipolar Bose gases and the BCS transition for dipolar Fermi
gases. In both cases we stress the dominant role of the trap geometry in
determining the properties of the system. We present also results concerning
bosonic dipolar gases in optical lattices and the possibility of obtaining
variety of different quantum phases in such case. Finally, we analyze
various possible routes towards achieving ultracold dipolar gases.
\end{abstract}

\pacs{PACS numbers: 03.75.Fi, 05.30.Jp, 64.60.Cn}

 }

\section{Introduction}

In the last twenty years at least three Nobel Prizes were awarded for
studies of the phenomena of condensation, superfluidity and
superconductivity \footnote{This paper is based on the lecture given by M. Lewenstein at
the Nobel Symposium ''Coherence and Condensation in Quantum Systems'',
Gothesburg, 4-7.12.2001.}, 
whereas several others were also related to
these subjects \cite{nobel}. The three Nobel Prizes in atomic, molecular and
optical (AMO) physics within the same period mark the path from trapping and
cooling of ions, through laser cooling of atoms, towards Bose-Einstein
condensation (BEC) of trapped atomic gases \cite{amo}. The observation of
BEC \cite{bec} has bursted a new interdisciplinary area of modern AMO and
condensed matter physics: the physics of ultracold {\it weakly interacting}
trapped quantum gases \cite{varenna}. So far, most of the experimental
results in this area can be very accurately modeled by mean-field
methods and its extensions, based on the Gross-Pitaevskii (GP) equation and
Bogoliubov-de Gennes equations for bosonic gases \cite{reviews}, and on the 
BCS theory for fermionic ones \cite{fermionsrev}.

This new research area presents several novel aspects, in particular the
finite character and inhomogeneity of the considered systems, and perhaps
more important, the unprecedented possibility of control and manipulation of
the system properties, which allows to study situations that were not yet
encountered in condensed matter or low temperature physics. The number of
parameters which can be successfully controlled is large, e.g. temperature,
number of atoms, trap potentials, interatomic interactions, etc. In
addition, the internal level structure of the atoms can also be employed,
e.g. to manipulate BECs using Raman-Bragg techniques \cite{bragg}, or to
analyze multicomponent condensates \cite{multi}

It is also worth mentioning, that the experimental techniques have progressed
recently to a stage at which mean field methods cease to provide a proper
physical picture. In this sense, recent experiments at JILA \cite{JILA} in
which the scattering length can be modified at will by using Feshbach
resonances, allow to realize systems with very large
scattering length, in which the mean-field picture is no more applicable.
Similarly, the realization of a Bose-condensed metastable Helium gas \cite
{He}, with a potential to study higher order correlation functions of the
system, demands an analysis beyond the mean field theory. The recent
demonstration of the Mott insulator to superfluid phase transition with
atoms in an optical lattice \cite{Greiner}, predicted in Ref. \cite{Jaksch},
belongs to the same category, but at the same time opens a new research area
of AMO physics: the physics of {\it strongly correlated} quantum gases. The
system in question allows for an easy and accurate control and manipulation,
and thus provides a new and particularly promising test ground for theories
of quantum phase transitions \cite{Sachdev}, which have traditionally dealt
with condensed-matter rather than atomic systems.

In recent years considerable interest has been devoted to another
aspect of the internal structure of the particles forming an ultracold gas,
namely their dipole moment. If such dipole moment is sufficiently large, the
resulting dipole-dipole forces may influence, or even completely change the
properties of BEC in bosonic gases, the conditions for BCS transition in
fermionic gases, or the phase diagram for quantum phase transitions for
ultracold dipolar gases in optical lattices. In this paper we present a
review of our recent results on ultracold dipolar gases.
We consider here only the case of polarized dipolar gases when all dipoles
are oriented in the same direction, and discuss the effects of the
dipole-dipole interactions and their interplay with short range (Van der
Waals) interactions. We do not consider several other important
aspects of dipole-dipole interactions, e.g. the role of the dipole-dipole
interactions in the spontaneous polarization of spinor condensates in
optical lattices \cite{Meystre}, or the self-bound structures in the field
of a traveling wave \cite{Giovanazzi}. As dipolar interactions can be quite
strong (relative to the typical low-energy collisions characterized by the $%
s$-wave scattering length), the dipolar particles are considered as
promising candidates for the implementation of fast and robust
quantum-computing schemes \cite{DDgates,qcomp}.

Our review is organized as follows. Section \ref{sec:bosons} 
is devoted to bosonic dipolar
gases, the prospects for their condensation, the ground-state properties of
the BEC, and its elementary excitations in a trap. In sections 
\ref{sec:fermions}, an
analogous analysis is applied to ultracold dipolar Fermi gases. In
particular, the possibility of the BCS transition is investigated. Section 
\ref{sec:lattices}
briefly considers the issue of quantum phase transitions in a dipolar Bose
gas placed in an optical lattice. In section \ref{sec:physical}
 we discuss various possible
physical systems which could lead to an ultracold dipolar gas. Finally, we
conclude in section \ref{sec:conclusions}.

\section{Ultracold dipolar Bose gases}
\label{sec:bosons}

In this section, we consider a system of $N$ bosonic particles possessing a
dipole moment, and confined in a harmonic trap of cylindrical symmetry. We
constraint our analysis to the case in which all dipoles are assumed to be
oriented along the symmetry axis of the trap. Additionally, we assume that the
particles interact via dipole-dipole forces, and that these forces either
play a dominant role, or at least compete with the short-range forces.

Why are {\em dipolar} gases interesting? The answer is simple: 
because they are {\em dipolar}. 
The dipole-dipole interaction potential between two dipolar
particles is given by 
\[
V_d({\bf R})= (d^2/R^3)(1-3\cos^2{\theta}), 
\]
where $d$ characterizes the dipole moment, ${\bf R}$ is the vector between the
dipoles ($R=|{\bf R}|$ being its length), and $\theta$ the angle between $%
{\bf R}$ and the dipole orientation (Fig. 1). The potential $V_d({\bf R})$
has two important properties: it is {\em anisotropic}, and is of {\em 
long-range} 
character. As we discuss below, these properties have important consequences.

At low temperatures one expects the dipolar Bose gases to condense. 
One expects also that the condensate properties will dramatically
depend on the geometry of the trap. In cigar-shape traps along the dipole
direction the interactions will be mainly attractive, and the condensate
will be unstable, similar to the case of a gas with attractive short-range 
interactions (negative $s$-wave scattering length) \cite{hulet}. 
Conversely, in pancake traps the interactions will be
mainly repulsive, and the gas might become stable. 
Therefore, the dipolar gases offer the unprecedented possibility 
of modifying the atom-atom interactions by turning an ``easy knob'', 
namely the trap geometry, which is relatively easy to control and modify 
experimentally. 

\begin{figure}[ht] 
\begin{center}\ 
\epsfxsize=7.0cm 
\hspace{0mm} 
\psfig{file=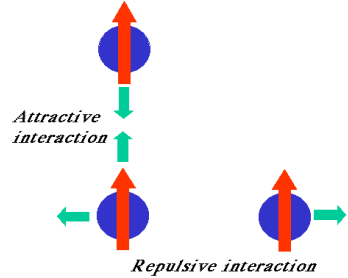,width=7.0cm}\\[0.1cm] 
\end{center} 
\caption{Anisotropy of dipole-dipole interactions.}
\label{dipdip}  
\end{figure}

Several groups have recently studied the physics of ultracold dipolar 
bosonic gases. L. You and his 
collaborators \cite{liyou,liyou2,recyou} have considered the
question of the validity of the mean-field approach in this case. This is a
non trivial question, since contrary to the case of short-range potentials, 
where the $s$-wave scattering always becomes dominant at very low 
temperatures, the scattering amplitude for dipole-dipole interactions 
has non vanishing contributions of all partial waves at low energies. 
Moreover, the interplay play between the dipole
interactions and the short range forces might lead to shape resonances in the
atom-atom scattering. Close to such resonances the effective scattering
length diverges, and the mean field approach cannot be used. Away from shape
resonances, however, the Gross-Pitaevskii (GP) equation, including the
non-local dipole-dipole interactions, provides a good description of the
condensate. Both L. You and his co-workers \cite{liyou}, G\'oral {\it et al.} 
\cite{goral} and Mackie {\it et al.} \cite{mackie} have investigated the
properties of the ground state of these systems, and the
interplay between dipolar forces and the short-range 
interactions. We have concentrated in our
studies \cite{santos} on the case of dominant dipole-dipole interactions.
This case is discussed in the next subsection.

\subsection{Ground state properties of dipolar Bose gases}

Similarly to \cite{liyou,goral}, we describe the dynamics of the condensate
wave function $\psi({\bf r},t)$ by using the time-dependent GPE: 
\begin{eqnarray}
&& i\hbar\frac{\partial}{\partial t}\psi({\bf r},t)= \left \{ -\frac{\hbar^2%
}{2m}\nabla^2+\frac{m}{2}(\omega_{\rho}^2\rho^2+\omega_z^2z^2)+ \right
\delimiter 0  \nonumber \\
&& \left \delimiter 0 + g|\psi({\bf r},t)|^2 +d^2\int d{{\bf r^{\prime}}} 
\frac{1-3\cos^2\theta} {|{\bf r}-{{\bf r^{\prime}}}|^3} |\psi({{\bf %
r^{\prime}}},t)|^2 \right \} \psi({\bf r},t).  \label{GPE}
\end{eqnarray}
Here $\psi({\bf r},t)$ is normalized to the total number of condensate
particles $N$. The third term in the rhs is the mean-field corresponding to
the short-range forces, whereas the last term is the mean field of
the dipole-dipole interaction. In the following, we omit the term $g|\psi(%
{\bf r},t)|^2\psi({\bf r},t)$, assuming that the interparticle interaction
is dominated by the dipole-dipole forces ($d^2\gg |g|=4\pi\hbar^2|a|/m$, where 
$a$ is the $s$-wave scattering length, and $m$ is the particle mass), and
that the system is away from shape resonances of $V_d({\bf R})$. The
ground-state properties are governed by the stationary GPE, in which the
lhs of Eq.\ (\ref{GPE}) is replaced by $\mu\psi({\bf r})$, where 
the chemical potential $\mu$ corresponds to the minimal energy solution.

There are two important parameters that can be easily 
controlled in experiments: 
the collective dipole strength $Nd^2$, and the trap aspect ratio 
$l=(\omega_\rho/\omega_z)^{1/2}=a_z/a_{\rho}$, where the characteristic
harmonic oscillator length $a_i=\sqrt{\hbar/m\omega_i}$. The first of these
parameters can be rewritten into a dimensionless form $\sigma=(Nd^2/a^3_{
{\rm max}})/\hbar\omega_{{\rm max}}$, where $a_{{\rm max}}= (\hbar/2M\omega_{
{\rm min}})^{1/2}$, with $\omega_{min}$ the minimum trap frequency.
Therefore, $\sigma$ represents the ratio of the dipole-dipole
interaction energy at the characteristic harmonic oscillator distance to the
characteristic harmonic oscillator energy, $\omega_{min}$.
Another quantity, which critically characterized the state of the 
system is the mean dipole-dipole interaction energy per particle given 
by the expression $V=(1/N)\int V_d( 
{\bf r}-{{\bf r^{\prime}}})\psi_0^2({\bf r})\psi_0^2({{\bf r^{\prime}}})d 
{\bf r} d{{\bf r^{\prime}}}$.

The results of Ref. \cite{santos} can be summarized as follows. For cigar
shaped traps with $l\geq 1$ the mean-field dipole-dipole interaction is
always attractive, and the gas becomes always unstable if the number of
particles $N$ exceeds a critical value $N_c$, which depends only on the trap
aspect ratio $l$. The quantity $|V|$ increases with $N$ and the shape of the
cloud changes. In spherical traps
the cloud becomes more elongated in the axial direction and near $N=N_c$ the
shape of the cloud is close to Gaussian with the cloud aspect ratio $%
L=L_z/L_{\rho}\simeq 2.1$. In cigar-shaped traps ($l\gg 1$) especially
interesting is the regime where $\hbar\omega_z\ll |V|\ll\hbar\omega_{\rho}$.
In this case the radial shape of the cloud remains the same Gaussian as in a
non-interacting gas, but the axial behavior of the condensate will be
governed by the dipole-dipole interaction which acquires a quasi
1-dimensional (1D) character. Thus, one has a (quasi) 1D gas with attractive
interparticle interactions, i.e. a stable (bright) soliton-like condensate,
where attractive forces are compensated by the kinetic energy \cite{soliton}%
. With increasing $N$, $L_z$ decreases. Near $N=N_c$, where $|V|$ is close
to $\hbar\omega_{\rho}$, the axial shape of the cloud also becomes Gaussian
and the aspect ratio takes the value $L\approx 3.0$.

The situation is quite different for pancake traps ($l\le 1$); in particular
there exists a critical aspect ratio $l_*\simeq 0.41$, which splits the pancake
traps into two different categories: soft pancake traps ($l_*\leq\l<1$) and
hard pancake traps ($l<l_*$). For soft pancake traps the dipole-dipole
interaction energy is positive for a small number of particles and increases
with $N$. The quantity $V$ reaches a maximum, and a further increase in $N$
reduces $V$ and makes the cloud less pancake. For a critical number of 
particles $N=N_c$ the BEC becomes unstable. 

We have found generally that the dipolar condensate is unstable and
collapses when $N>N_c$ for $V<0$ with $|V|>\hbar\omega_{\rho}$. However, for
hard pancake traps with $l<l_*$ the condensate is stable at any $N$, because 
$V$ always remains positive. For small $N$ the shape
of the cloud is Gaussian in all directions. With increasing $N$, the
quantity $V$ increases and the cloud first becomes Thomas-Fermi in the
radial direction and then, for a very large $N$, also axially. The ratio of
the axial to radial size of the cloud, $L=L_z/L_{\rho}$, continuously
decreases with increasing number of particles and reaches a limiting value
at $N\rightarrow\infty$. In this respect,
for a very large $N$ we have a pancake Thomas-Fermi condensate.

It is worth stressing that many of the above results, calculated from a 
direct numerical simulation of the time-independent GPE, were also 
analytically obtained with the help of a variational method already used 
in the context of short-range interacting BEC (c.f. \cite{baym,negat}). 

\begin{figure}[ht] 
\begin{center}\ 
\epsfxsize=7.0cm 
\hspace{0mm} 
\psfig{file=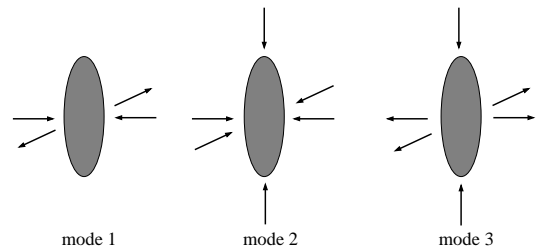,width=7.0cm}\\[0.1cm] 
\end{center} 
\caption{Graphical representation of oscillations modes of the condensate.}
\label{modes}  
\end{figure}

\subsection{Excitations in dipolar Bose gases}

In this section we analyze the elementary excitations in a trapped dipolar 
gas. One possible method of analysis of these excitations could be 
to solve the corresponding Bogoliubov-de Gennes equations \cite{reviews},
which become non-local due to the presence of dipole-dipole interactions .
Another approach, suggested in the context of short-range interacting 
condensates by Ruprecht {\it et al.} \cite{Ruprecht}, 
is to use the time-dependent GPE, and numerically study the spectrum 
of the small perturbations around the ground-state solution. 
For the lowest lying excitation modes, such those schematically
sketched in Fig. \ref{modes}, it is also possible to obtain analytic results
using the time dependent variational method of Ref. \cite{Perez}, which 
provides very accurate results when $\sigma$ is not to close to its critical 
value $\sigma_c$, which is defined as the dipole strength at which the 
BEC becomes unstable. L. You and his collaborators \cite{liyou2,recyou} 
have recently employed this method to analyze various properties of 
the excitations in dipolar Bose gases. We have recently \cite{werec} 
used the numerical method of Ref.\ \cite{Ruprecht} and the variational 
approach of Ref. \cite{Perez}, in order to answer two fundamental questions
concerning elementary excitations, namely how is the qualitative nature of
the instability, and how the effects of the dipole-dipole interactions 
can be observable from the excitation spectrum.

The question of the nature of instability is twofold. First, the 
determination of which of the modes becomes unstable when the number 
of atoms reaches the critical value. Second, the analysis of the behavior 
of the corresponding mode frequency close to the criticality. This analysis 
has been first performed for the case of a Bose gas with attractive 
short-range interactions. Bergeman \cite{tom} observed from his 
numerical results that, as the ratio of the nonlinear interaction energy 
to the trap frequency, $\gamma$, approaches the critical value, 
the frequency of the ``breathing'' mode $2$ tends to zero and merges with 
the frequency of the Goldstone mode corresponding to the overall phase 
of the condensate. The parameter $\gamma$ is defined as 
\begin{equation}
\gamma =\sqrt{\frac{2}{\pi}}\frac{a}{\bar\sigma_0}N,
\end{equation}
where $\bar\sigma_0$ is the geometric mean width of the ground state in an
ellipsoidal harmonic trap. In the interesting region of $a<0$, 
the critical value of $\gamma $ is $\gamma_c=-0.54$. 
Above criticality, the ``breathing''
becomes unstable and attains complex frequency. Singh and Rokhsar analyzed
this instability \cite{exc_th} using self-similar solutions to describe the
modes, or equivalently the variational approach of the previous subsection. 
They have shown that close to criticality the frequency of the mode 
$2$ vanishes as $|\gamma -\gamma_c|^{1/4}$.

In the case of dipolar gases with dominant dipole interactions the situation
is completely different and much more complex. Only for aspect ratios far
above the criticality, $l\gg l^*$ ($l>1.29$) the situation reminds that of a
gas with attractive short-range interactions. The mode corresponding to the
lowest frequency is the "breathing" mode $2$. This mode becomes unstable
when the parameter $\sigma\to \sigma_c$. The scaling behavior of its
frequency $\omega_2$ can be analyzed employing the variational approach of
the previous section, and the analytic (although approximate) 
expressions for $\omega_2$ from Ref. \cite{liyou2}. 
We find that $\omega_2$ goes to zero as 
$(\sigma_c-\sigma)^\beta$, with $\beta\simeq 1/4$.

For intermediate values of $l>l_*$ ($0.75<l<1.29$), the exponent $\beta$ is
still close to 1/4, but the "breathing" and quadrupole modes mix. For 
$\sigma$ far below $\sigma_c$ the mode corresponding to the lowest frequency
is the "breathing" mode $2$. As we approach the critical value of $\sigma$
the character of the lowest frequency mode changes and becomes
quadrupole-like.

\begin{figure}[ht] 
\begin{center}\ 
\epsfxsize=6.0cm 
\hspace{0mm} 
\psfig{file=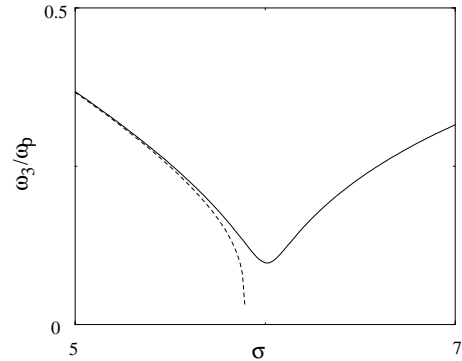,width=6.0cm}\\[0.1cm] 
\end{center} 
\caption{Frequency $\omega_3$ of the lowest mode 3 as a function of $\sigma$
for $l$ just below (solid line), and just above (dashed line) $l^*$. For $%
l<l^*$, the frequency $\omega_3$ has a quadratic minimum close to $%
\sigma=\sigma_c(l^*)$.} 
\label{stabl} 
\end{figure}

For $l$ close to $l^*$ ($l<0.75$) the situation changes and the mode
corresponding to the lowest frequency is the quadrupole mode $3$. Now, it is
its frequency $\omega_3$ which tends to zero as the parameter $\sigma$
approaches the critical value. For $l$ not too close to $l^*$ the exponent 
$\beta$ is still close to $1/4$. Completely new effects arise due to the
existence of the previously discussed critical aspect ratio 
$l^*\simeq 0.41$. This is illustrated in Fig. 
\ref{stabl}. As one approaches $l^*$, the exponent $\beta$ departs from $1/4$
towards a greater value. This crossover is explained in Fig. \ref{stabl}.
For $l$ slightly below $l^*$, the frequency of the quadrupole mode $\omega_3$
has a quadratic minimum close $\sigma=\sigma_c(l^*)$. Exactly at $l=l^*$, $%
\omega_3 $ goes thus to zero as $(\sigma_c-\sigma)^2$, i.e. $\beta=2$.

\begin{figure}[ht] 
\begin{center}\ 
\epsfxsize=7.0cm 
\hspace{0mm} 
\psfig{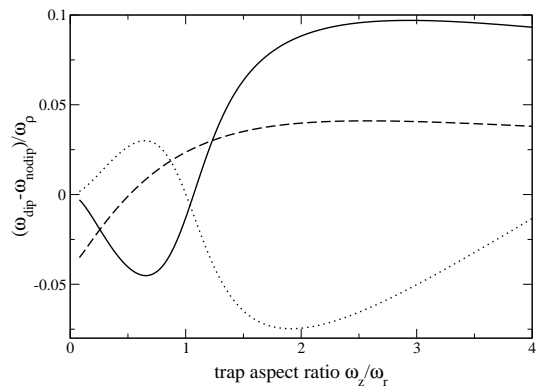}\\[0.1cm] 
\end{center} 
\caption{Difference of excitation frequency for modes 1 (dashed line), 2
(dotted line) and 3 (solid line) between the cases of purely contact and
mixed (contact and dipole-dipole) interactions as a function of the trap
aspect ratio for 10000 atoms and $a=a_{{\rm Na}}$ (data from variational
analysis).} 
\label{fig_vsasp} 
\end{figure}

Apart from the question of the nature of instability, in Ref. \cite{werec}
we have considered also the situation in which the particles interact via
both contact and dipole-dipole forces. In this case the strength for both
types of forces is comparable and neither of them can be neglected. As a
specific example we have considered the case of $^{52}$Cr, which has drawn
some experimental interest \cite{Weinstein,Pfau1,Celotta,Pfau2}. Chromium
has a large magnetic dipole moment of $6 \mu_B$ (Bohr magnetons), but its $s$
-wave scattering length is still unknown. We have considered the trap
frequencies of an experiment underway at the University of Stuttgart \cite
{Pfaupriv}: $\omega_z=2 \pi 40$ Hz and $\omega_{\rho}=2 \pi 485$ Hz. Our
results suggest clearly that the effects of dipole-dipole interaction would
be accessible to experimental detection if condensation of chromium or
europium \cite{europium} is achieved, and if their scattering lengths 
are not anomalously large. Dipole-dipole effects can be enhanced
by reducing the $s$-wave scattering by employing Feshbach resonances \cite
{FeshbachMIT,85Rb}. It is thus very important to analyze possible parameters
(scattering length, number of atoms, trap aspect ratio) that would maximize
the predicted frequency shifts. A typical results of our analysis is shown
in Fig. \ref{fig_vsasp}.

\section{Dipolar Fermi gases}
\label{sec:fermions}

The recent success in observing the quantum degeneracy in ultra-cold atomic
Fermi gases \cite{coolfer1,coolfer2,coolfer3,coolfer4} stimulates a search
for gaseous Fermi systems where combined effects of Fermi statistics and
interparticle interactions result in a non-trivial physical behavior. Due to
the Pauli principle, to observe these effects in the case of a short-range
Van der Waals interparticle interaction, a simultaneous trapping of at least
two different fermionic species is needed, with a rather severe constraint
on their relative concentrations. The situation is different for Fermi gases
of dipolar particles.  In the ultra-cold limit, the
dipole-dipole scattering amplitude is energy independent for any orbital
angular momenta $L>0$. This observation follows from the results of 
\cite{orbital,verhaar}. It has also been recently discussed with important 
physical insight and consequences by You and coworkers 
\cite{you-marinescu0,deb-you}. This opens
prospects to observe the effects of interparticle interaction in a {\it %
single-component} Fermi gas, where only scattering with odd orbital momenta
(negligible in the case of Van der Waals interactions) is present. These
prospects are especially interesting as in single-component fermionic gases
the Pauli exclusion principle provides a strong suppression of inelastic
collisional rates (see \cite{verhaar}). Hence, one can think of achieving
significantly higher densities than in Bose gases.

\subsection{Physical system}

We consider a single component gas of fermionic dipolar particles in a
harmonic trap, with dipole moments oriented in the same direction ($z$
-axis). Then, due to the Pauli principle, the contribution of the
short-range part of the interparticle interaction can be neglected, and,
therefore, the dominant interaction between particles is the dipole-dipole
one. As a result, the Hamiltonian of the system reads 
\begin{eqnarray}
H &=&\int d{\bf r}\hat\psi ^{\dagger }({\bf r})\left\{ -\frac{\hbar ^{2}}{2m}
\Delta +V_{{\rm trap}}({\bf r})-\mu \right\} \hat\psi ({\bf r})
\label{hamiltonian} \\
&&+\frac{1}{2}\int d{\bf r}d{\bf r}^{\prime }\hat\psi ^{\dagger }({\bf r})
\hat\psi (
{\bf r})V_{{\rm d}}({\bf r-r}^{\prime })\hat\psi ^{\dagger }({\bf r}^{\prime
})\hat\psi ({\bf r}^{\prime }),
\end{eqnarray}
where $V_{{\rm d}}({\bf r})=(d^{2}/r^{3})(1-3\cos ^{2}(\theta _{{\bf r}}))$
is the dipole-dipole interaction with $\theta _{{\bf r}}$ being the angle
between the interparticle distance ${\bf r}$ and the $z$-axis, $\mu $ is the
chemical potential, and $V_{{\rm trap}}({\bf r})=m[\omega
_{z}^{2}z^{2}+\omega _{\rho }^{2}(x^{2}+y^{2})]/2$ is the trapping potential; 
$\hat\psi(\bf r)$ and $\hat\psi^\dag(\bf r)$ denote annihilation and 
creation operators for fermions with canonical anticommutation relations.

We assume that the gas is in the regime of quantum degeneracy, i.e. the
temperature $T$ is much smaller than the chemical potential $\mu $ (or the
Fermi energy $\varepsilon _{F}$), $T\ll \mu =\varepsilon _{F}=(\hbar
^{2}/2m)(6\pi ^{2}n_{0})^{2/3}$, where $n_{0}$ is the maximal gas density
(in the center of the trap), and the interparticle interaction is weak. The
latter means that the mean dipole-dipole interaction energy per particle $%
nd^{2}$ is much less than the Fermi energy, $nd^{2}\ll \varepsilon _{F}$.

\subsection{Why are dipolar Fermi gases interesting?}

Which features make this system attractive and challenging for both
theorists and experimentalists? The system under consideration is unique in
that the physics is dominated by the long-range dipole-dipole forces (the
short-range part of the interparticle interaction, as already mentioned, can
be neglected due to the fermionic nature of the particles). The
dipole-dipole interaction is anisotropic and, as a result, it mixes
scattering channels with different angular momenta $l$. Additionally, the
interaction is partially repulsive (when two particles are side by side to
each other) and partially attractive (when they are on top of each other).
This means that, similarly as for dipolar Bose gases, 
the properties of the system depend on the trap geometry
(see Ref.\cite{dipoleWKB}). Indeed, in a pancake trap, most of the time the
particles are side by side to each other, and the average interparticle
interaction is repulsive. In the opposite case of a cigar shaped trap, the
particles are mostly on top of each other and, as a result, the average
interaction is attractive. Therefore, one can expect a pronounced dependence
of the system properties on the aspect ratio 
$l=\sqrt{\omega_{\rho}/\omega_{z}}$ of the trap.

The fact that the dipole-dipole interparticle interaction is partially
attractive opens the possibility for the BCS pairing at sufficiently low
temperatures. For example, the matrix element $\left\langle L=1,M=0\left| 
V_{{\rm 
d}}\right| L=1,M=0\right\rangle $ of the dipole-dipole interaction between
states with angular momentum $L=1$ and its projection to the $z$-axis $M=0$,
is negative: 
\[
\left\langle L=1,M=0\left| V_{{\rm d}}\right| L=1,M=0\right\rangle =-\frac{
4\pi }{5}d^{2}<0. 
\]
This signals about the possible BCS transition 
\cite{dipstoof,you-marinescu,dipoleTc}.

\subsection{The BCS pairing}

The BCS pairing transition corresponds to the formation of Cooper pairs, the
correlated states of two fermions. The corresponding order parameter $\Delta 
$ that appears below the transition temperature $T_{c}$, describes the
coherent motion of the Cooper pairs: 
\[
\Delta ({\bf r}_{1},{\bf r}_{2})\propto \left\langle \psi ({\bf r}_{1})\psi
( {\bf r}_{2})\right\rangle \neq 0. 
\]

The order parameter vanishes for $T\rightarrow T_{c}$ and obeys the gap
equation. For temperatures close to $T_{c}$ this equation has the form (the
so-called linearized gap equation): 
\begin{equation}
\Delta ({\bf r}_{1}{\bf ,r}_{2})=-V_{{\rm d}}({\bf r}_{1}{\bf -r}_{2}{\bf )}%
\int d{\bf r}_{3}d{\bf r}_{4}K({\bf r}_{1}{\bf ,r}_{2}{\bf ;r}_{3}{\bf ,r}%
_{4})\Delta ({\bf r}_{3}{\bf ,r}_{4}),  \label{lingapeq}
\end{equation}
where the kernel $K$ is defined as 
\begin{eqnarray*}
K({\bf r}_{1}{\bf ,r}_{2}{\bf ;r}_{3}{\bf ,r}_{4}) &=&\sum_{\nu _{1},\nu
_{2}}\frac{\tanh (\xi _{\nu _{1}}/2T)+\tanh (\xi _{\nu _{2}}/2T)}{\xi _{\nu
_{1}}+\xi _{\nu _{1}}} \\
&&\phi _{\nu _{1}}({\bf r}_{1})\phi _{\nu _{2}}({\bf r}_{2})\phi _{\nu
_{1}}^{*}({\bf r}_{3})\phi _{\nu _{2}}^{*}({\bf r}_{4})
\end{eqnarray*}
with $\xi _{\nu }=\varepsilon _{\nu }-\mu $ and $\phi _{\nu }({\bf r})$
being the solutions of the single-particle Schr\"{o}dinger equation 
\[
\left\{ -\frac{\hbar ^{2}}{2m}\Delta +V_{{\rm trap}}({\bf r)}\right\} \phi
_{\nu }({\bf r})=\varepsilon _{\nu }\phi _{\nu }({\bf r}). 
\]

The critical temperature $T_{c}$ can be defined as the maximum temperature
for which the linearized gap equation (\ref{lingapeq}) has a nontrivial
solution. It should be mentioned that this equation can only be used for
finding the critical temperature and the spatial dependence of the order
parameter. To determine the absolute value of the order parameter, which is
temperature dependent, one has to consider the terms nonlinear in $\Delta $,
omitted in Eq.(\ref{lingapeq}).

\subsection{Spatially homogeneous dipolar Fermi gases}

In the spatially homogeneous case ($V_{{\rm trap}}({\bf r)=}0$), the order
parameter depends only on the relative coordinate ${\bf r=r}_{1}-{\bf r}_{2}$%
, $\Delta ({\bf r}_{1}{\bf ,r}_{2})=\Delta ({\bf r}_{1}{\bf -r}_{2})$, and
the gap equation (\ref{lingapeq}) with the leading non-linear term added,
was discussed in Ref.\ \cite{dipoleTc},  
where it was found that the critical temperature of the
BCS transition in the spatially homogeneous Fermi gas of dipolar particles
equals (including the so-called Gor'kov and Melik-Barkhudarov corrections)
\begin{equation}
T_{c}=1.44\varepsilon _{F}\exp (-\frac{\pi \varepsilon _{F}}{12nd^{2}}%
)=1.44\varepsilon _{F}\exp (-\frac{\pi \hbar }{2\left| a_{d}\right| p_{F}}),
\label{Tc}
\end{equation}
where $n$ is the gas density, $\varepsilon _{F}=p_{F}^{2}/2m=(\hbar
^{2}/2m)(6\pi ^{2}n)^{2/3}$ the Fermi energy, $p_{F}$ the Fermi momentum,
and $a_{d}=-2md^{2}/\pi ^{2}\hbar ^{2}$ the effective scattering length. The
latter has been introduced in order to make the exponent in the expression
for $T_{c}$, Eq.(\ref{Tc}), look like the exponent in the expression for the
critical temperature in the case of a two-component Fermi gas with the
inter-component $s$-wave scattering length $a_{d}$. For the case of a
single-component gas of fermionic ND$_{3}$ molecules with the dipole moment $%
d=1.5$D, the transition temperature $T_{c}$ is larger than $100$ nK at
densities $n>5\cdot 10^{12}$ cm$^{-3}$.

The properties of the superfluid dipolar Fermi gas are different from those
of a two-component fermionic gas with $s$-wave pairing due to a short-range
inter-component interaction. In the case of the $s$-wave paring, the order
parameter is isotropic, whereas it is anisotropic in the superfluid dipolar
Fermi gas. In the latter case, the value of the order parameter $\Delta $ at
the Fermi surface $p=p_{F}$, is

\[
\Delta (p_{F},\theta )=2.5T_{c}\sqrt{(T_{c}-T)/T_{c}}\cdot \sqrt{2}\sin (%
\frac{\pi }{2}\cos (\theta )), 
\]
where $\theta $ is the angle between the momentum ${\bf p}$ and the $%
z $-axis. As a result, the gap in the spectrum of single-particle
excitations, which appears below the transition temperature $T_{c}$, is
anisotropic. For example, excitations with momenta in the direction of the
dipoles acquire the largest gap. In contrast to this, the eigenenergies of
excitations with momenta perpendicular to the dipoles remain unchanged. The
properties of collective excitations are also expected to be dependent on
the direction of their momenta. Therefore, the response of the dipolar
superfluid Fermi gas to small external perturbations will have a pronounced
anisotropic character.

Another distinguishing feature of the superfluid dipolar Fermi gas is
related to the temperature dependence of the specific heat. Well below the
critical temperature the single-particle contribution to the specific heat
is proportional to $T^{2}$, rather than being exponentially small as in the
case of the $s$-wave pairing. This follows from the fact that the energy $%
\varepsilon $ of single-particle excitations has a line of zeros on the
Fermi surface: $\varepsilon (p_{F})=0$ for the angles at which $\Delta
(p_{F},{\bf n})=0$, i.e. for $\theta =\pi /2$ and an arbitrary azimuthal
angle $\varphi $. As a consequence, the density of states in the vicinity of
the Fermi energy is $\nu (\varepsilon )\sim \varepsilon $ for $\varepsilon
\ll \Delta _{0}$. Therefore, at temperatures $T\ll \Delta _{0}\sim T_{c}$,
the temperature dependent part of the energy of the system is proportional
to $T^{3}$, and the specific heat is thus proportional to $T^{2}$. This
contribution is much larger than that of collective modes, which is $\propto
T^{3}$ and is dominant in the case of the $s$-wave pairing.

\subsection{BCS pairing in a harmonic trap}

We now discuss how the results of the previous Section change when the
harmonic trapping potential is switched on. It turns out (see Ref.\cite
{ferdiptrap} for details), that the presence of the trapping potential
always results in the decrease of the critical temperature of the superfluid
transition. In the case where the trap frequencies $\omega _{z}$, $\omega
_{\rho }$ are much smaller than the critical temperature $T_{c}$, Eq.(\ref
{Tc}), of the transition in the spatially homogeneous gas with density $n$
equal to the maximum gas density $n_{0}$ in the trap, one obtains 
\begin{eqnarray}
\frac{T_{c}^{{\rm trap}}-T_{c}}{T_{c}} &=&-\frac{\omega }{T_{c}}\sqrt{\frac{%
7\zeta (3)}{48\pi ^{2}}\left( 1+\frac{\pi \varepsilon _{F}}{24nd^{2}}\right) 
}  \label{Tctrap} \nonumber \\
&&\left\{ 2\sqrt{1-\frac{3}{\pi ^{3}}}l^{2/3}+\sqrt{1+\frac{6}{\pi
^{3}}}l^{-4/3}\right\} ,
\end{eqnarray}
where $T_{c}^{{\rm trap}}$ is the transition temperature in the trap, $\zeta
(z)$ the Riemann zeta-function, and $\omega =(\omega _{z}\omega _{\rho
}^{2})^{1/3}$. 
One can see from this expression that for sufficiently small $%
\omega /T_{c}$, the critical temperature in the trap $T_{c}^{{\rm trap}}$ is
only slightly lower than $T_{c}$. As follows from Eq.(\ref{Tctrap}), the
optimal value of the trap aspect ratio $l$, that corresponds to
maximal $T_{c}^{{\rm trap}}$ at a given $T_{c}$, is $l\simeq 1.38$.

Another interesting feature of the BCS transition in a trapped Fermi gas of
dipolar particles is the existence of the critical value for $\omega _{z}$.
This comes from the fact that the paired states in the trapped dipolar Fermi
gas have different quantum number $n_{z}$, and, therefore, their energies
are always different, at least by the amount $\omega _{z}$. When the
difference becomes of the order of the order parameter $\Delta \sim T_{c}$
which measures the strength of the paring correlations, the pairing is
obviously impossible. As a result, the superfluid transition in the dipolar
Fermi gas is possible only in traps with $\omega _{z}<\omega _{zc}$, where
the critical frequency $\omega _{zc}$ is found to be $\omega _{zc}=1.8T_{c}$%
. As can be seen from Eq.(\ref{Tctrap}), confinement in the radial direction
decreases the critical temperature as well. Therefore, in general one would
expect the existence of the critical aspect ratio $l{c}$ such that
the pairing is possible only in a trap with $l>l _{c}$.

\section{Dipolar Bose gases in optical lattices}
\label{sec:lattices}

In a very recent paper \cite{lattice}, we have investigated the
ground-state properties of a polarized gas of bosonic particles (atoms or
molecules) possessing a dipole moment and placed in an optical lattice. This
system presents features which are novel both in the theory of quantum phase
transitions, and in the context of degenerate quantum gases. Recently,
Jaksch {\it et al.} \cite{Jaksch} analyzed the superfluid-Mott insulator phase
transition in the context of cold bosonic atoms with short-range interactions 
in an optical lattice. 
This analysis has been very recently confirmed experimentally 
\cite{Greiner}. The dipole-dipole interactions, which are anisotropic and
have a long range, have not been studied, to the best of our knowledge, in
the context of the Mott insulator to superfluid phase transition. In
addition, we shall show that the long-range character of the dipole-dipole
interaction allows for the existence of not only Mott-insulating and
superfluid phases, but also several other phases in the system. Very
importantly, the gas of cold dipolar bosons in an optical lattice is shown
to be a system with easily tunable interactions which may permit to realize
all the different phases experimentally. The availability of such a highly
controllable system may be crucial in answering some unresolved questions
and controversies in the theory of quantum phase transitions (e.g. the
existence of a yet unobserved supersolid \cite{Leggett}, or a Bose metal at
zero temperature \cite{metal}).

A dilute gas of bosons in a periodic potential (e.g. in an optical lattice)
can be described with the help of the Bose-Hubbard model \cite{Jaksch}. For
particles interacting via long-range forces the Bose-Hubbard Hamiltonian
becomes:

\begin{eqnarray}  \label{H}
H&=&J\sum_{<i,j>}b_{i}^{\dagger}b_{j} +\frac{1}{2}U_{0}\sum_{i}n_{i}(n_{i}-1)
\nonumber \\
&+&\frac{1}{2}U_{1}\sum_{<i,j>}n_{i}n_{j}+\frac{1}{2}U_{2}%
\sum_{<<i,j>>}n_{i}n_{j} +\ldots,
\end{eqnarray}

\noindent where $b_{i}$ is an operator annihilating a particle at a lattice
site $i$ in a state described by the Wannier function $w({\bf r}-{\bf r}%
_{i}) $ of the lowest energy band, localized on this site. ${\bf r}_{i}$ is
the position of the local minimum of the optical potential, and $%
n_{i}=b_{i}^{\dagger}b_{i}$ is the number operator for the site $i$. In Eq. (%
\ref{H}) only the nearest neighbor tunneling is considered, which is
described by the parameter $J$. The interparticle interactions are
characterized by the parameters $U_m$, where $m=|j-i|$. In particular, $U_0$
determines the on-site interactions, $U_1$ the nearest-neighbor
interactions, $U_2$ - the interaction between the next-nearest neighbors,
etc. Consequently, the respective summations in Eq.(\ref{H}) must be carried
out over appropriate pairs of sites which are marked by $<>$ for the nearest
neighbors, $<<>>$ for the next-nearest neighbors, etc. 
Note that in 2D and 3D, $i$ is a multi-index enumerating 
sites in the corresponding lattice.

If only short-range interactions are present, seminal theoretical studies
of the Bose-Hubbard model and related theory of arrays of Josephson 
junctions \cite{Fisher}, show that two different quantum phases can 
occur for bosons in a lattice, either superfluid or Mott-insulator.
In the case of finite-range interactions additional phases as supersolid and 
checker-board are expected \cite{schon}.

We have found the ground state of the system by employing a 
variational approach based in the so-called Gutzwiller Ansatz 
(see \cite{Jaksch} and references therein). We have shown that by modifying
well-controllable parameters a whole variety of different quantum phases can be
achieved, including superfluid, supersolid, Mott-insulator, checker-board
and collapse phases. This possibility of manipulation of the corresponding 
quantum phases by modifying easily controllable external parameters (e.g. 
the on-site aspect ratio) makes the dipolar Bose gas in an optical 
lattice particularly challenging for theory and experiments. 
In particular, possible applications for quantum
information processing are worth mentioning in this context \cite
{DDgates,qcomp}.

\section{Physical realizations of ultracold dipolar gases}
\label{sec:physical}

One of the possible physical realizations of a gas of dipolar particles 
is provided by electrically polarized gases of polar molecules. 
This molecules can have a large permanent electric dipole. 
The creation of cold clouds of polar molecules has been recently
demonstrated in experiments with buffer-gas cooling \cite{buffercool}, 
as well as in experiments based on deceleration and cooling of polar 
molecules by time-dependent electric fields \cite{movefield}.
The molecular dipole moments typically range from $0.1$ D 
($1$ D$=10^{-18}{\rm charge\  SGS\times cm}$) to $%
1 $ D. For bosonic molecules that should be sufficient for achieving BEC at
relatively low density of the gas. For example, the dipole moment of the
fermionic deuterated ammonia molecule $^{15}$ND$_{3}$ is $d=1.5$ D, which
corresponds to an effective scattering length (see above, Eq.(\ref{Tc})) of $%
a_{{\rm d}}=-1450$ \AA. This is even larger than the scattering length for
the inter-component interaction in the widely discussed case of fermionic $%
^{6}$Li. On the other hand, for the fermionic molecule $^{14}$N$^{16}$O with
dipole moment $d=0.16$ D, the corresponding scattering length $a_{{\rm d}%
}=-24$ \AA.

Another possibility is to use magnetic atomic dipoles 
\cite{Weinstein,Pfau1,Celotta,Pfau2,Pfaupriv,europium}. 
Chromium atoms
have a magnetic moment $\mu =6\mu _{B}$, which is equivalent to an electric
dipole moment of $d^{*}=6\cdot 10^{-2}$ D, and an effective scattering
length $a_{{\rm d}}=-5$ \AA . This is too small and is unlikely to result in
any interesting fermionic effects, such as BCS pairing. As we mentioned in
section \ref{sec:bosons}, however, 
this could be enough to observe the influence of
dipole-dipole interactions on the elementary excitations of Chromium
BEC, provided the (still unknown) $s$-wave scattering length 
is not anomalously large, 
and it takes a value in the range of few tens of \AA.

Permanent electric 
dipole moments can also be created by applying a high dc electric
field to an atom. This possibility was discussed in Ref.\ \cite
{you-marinescu0}, and here we only mention that in order to induce the
dipole moment of the order of $0.1$ D (the corresponding scattering length $%
a_{d}\sim -10\div 100$ \AA ) one needs an electric field of the order of $%
10^{6}$ V/cm. Nevertheless, the influence of dipole forces on elementary
excitation spectrum of a dipolar BEC might be in this case also perhaps
observable using not as high electric fields.

Finally, we mention the possibility of inducing a time averaged electric dipole
moment of an atom by a stroboscopic laser coupling of the ground atomic
state to a Rydberg state in a moderate dc electric field. 
This method has been proposed in the
context of bosonic gases in Ref. \cite{santos}. 
We propose to place a BEC of alkali atoms into a moderate static 
electric field. The idea is to admix, with the help of a laser, to 
the atomic ground state the permanent dipole moment of a low-lying 
Rydberg state. 
Rydberg states of Hydrogen and alkali atoms exhibit a linear Stark
effect \cite{Gallagher}: in Hydrogen, for example, an electric field $E_{s}$
splits the manifold of a Rydberg states of given principal quantum number $n$
and magnetic quantum number $m$ into $2(n-|m|-1)$ Stark states. The
outermost Stark states have (large) permanent dipole moments $d_{R}\sim
n^{2}ea_{B}$ (with $a_{B}$ the Bohr radius), and there will be an associated
dipole-dipole force between atoms. The spacing $\hbar \omega _{s}\simeq
nea_{B}E_{s}$ between adjacent Stark states should greatly exceed the
mean-field dipole-dipole interaction (and the gas temperature) in order to
avoid interaction-induced transitions from the lowest sublevel to other
sublevels of the manifold.

This dipole-dipole interaction can be controlled with a laser \cite{DDgates}%
. This is achieved either by admixing the permanent dipole moment of the
Stark states to the atomic ground state with an off-resonant cw laser, or by
a stroboscopic excitation with a sequence of laser pulses. The pulses should
be separated by the time $T$, have duration $2\Delta t\ll T$ and area
multiple of $2\pi$.

The field $E_s$ and the laser should be chosen such that they do
not couple the selected lowest sublevel to other Rydberg (sub)states.
The stroboscopic excitation ``dresses'' the atomic internal states, 
so that each
atom acquires a time averaged dipole moment of the order of $d_s=n^2ea_B f$,
oriented in the direction of $E_s$, where $f=\Delta t/T$. 
Even though the quantity $f$
is assumed to be small, the induced dipole can be rather large for $n\gg 1$.
Taking for example $\Delta t=1$ns, $T=10\mu$s, and $n=20$, we obtain $%
d_s=0.1 $D. The 
resulting time dependent Hamiltonian can be replaced by its time average,
leading to Eq.(\ref{GPE}) with $d=d_s$. The characteristic time scale 
in Eq.(\ref{GPE}) is provided by the inverse of the trap frequency 
$\omega^{-1}$. Hence, in our case the dynamics of the system is described 
by Eq.(\ref{GPE}) with $d=d_s$, if the condition $\Delta t,T\ll\omega^{-1}$ 
is satisfied.

It is important to note that, for the values of the scattering length of the
order of $-10\div 100$ \AA ,\ one nevertheless has a possibility to achieve
the critical temperature of the BCS transition of the order of $100$ nK at
densities $n\sim 10^{16}$ cm$^{-3}$. In the case of a single-component
atomic Fermi gas, such densities are not unrealistic because the inelastic
processes, such as two-body collisions and three-body recombination, that
usually limit the maximum value of the gas density, are strongly suppressed. 
This
comes from the fact that these processes take place at short interatomic
distances (of the order of tens of Angstroms) where the wave function of the
relative motion of two identical fermions vanishes due to the Pauli
principle.

\section{Conclusions}
\label{sec:conclusions}

In this paper we have reviewed the properties of the bosonic and fermionic
dipolar gases, and analyzed the perspectives that these systems can offer.
We have shown, that the physics of these systems can differ qualitatively in
a significant way with respect to the gases interacting via Van der Waals
forces. In addition, new easy ways of control and manipulation are possible 
for dipolar gases. For all of that, we consider that the
ultracold dipolar quantum gases constitute an interesting challenge for
theory and experiments in AMO and condensed-matter physics.

We acknowledge support from the Alexander von Humboldt Stiftung, the
Deutscher Akademischer Austauschdienst (DAAD), the Deutsche
Forschungsgemeinschaft, the European Union Network "Coherent Matter Wave
Interactions", the ESF Program BEC2000+,  the Russian Foundation 
for Fundamental Studies, 
the Polish KBN (grant no 5 P03B 102 20) and from the subsidy
of the Foundation for Polish Science. We thank B. Altschuler, 
I. Bloch, M. Greiner, G. Meijer, J. Martikainen, K. Rz\c a\.zewski, 
G. Sch\"on, G. V. Shlyapnikov,  and P. Zoller for very entlighting discussions.

\end{document}